\documentclass[
  prl,
  twocolumn,
  superscriptaddress,
  aps,
  amsmath,
  amssymb,
  floatfix,
  10pt,
  longbibliography
]{revtex4-2}

\usepackage{graphicx}
\usepackage{bm}
\usepackage{mathtools}
\usepackage{microtype}
\usepackage{ragged2e}
\usepackage{xcolor}
\usepackage[colorlinks=true,allcolors=blue]{hyperref}

\newcommand{\ii}{\mathrm{i}}
\newcommand{\ee}{\mathrm{e}}
\newcommand{\US}{U_{SS}}
\newcommand{\Uc}{U_c}
\newcommand{\Heff}{H_{\mathrm{eff}}}
\newcommand{\Tr}{\operatorname{Tr}}
\newcounter{appref}

\begin{document}

\title{Emergent Non-Biorthogonal Born Rule at Dynamical Quantum Phase Transitions}

\author{Wei-Wei Yang}
\affiliation{Asia Pacific Center for Theoretical Physics, Pohang, Republic of Korea}
\affiliation{Beijing National Laboratory for Condensed Matter Physics and Institute of Physics, Chinese Academy of Sciences, Beijing 100190, China}

\author{Henning Schomerus}
\email[Contact author: ]{h.schomerus@lancaster.ac.uk}
\affiliation{School of Physics and Astronomy, Lancaster University, Lancaster, LA1 4YB, United Kingdom}

\author{Jiangping Hu}
\email[Corresponding author: ]{jphu@iphy.ac.cn}
\affiliation{Beijing National Laboratory for Condensed Matter Physics and Institute of Physics, Chinese Academy of Sciences, Beijing 100190, China}
\affiliation{New Cornerstone Science Laboratory, Beijing, 100190, China}

\author{Zhesen Yang}
\email[Corresponding author: ]{yangzs@xmu.edu.cn}
\affiliation{Asia Pacific Center for Theoretical Physics, Pohang, Republic of Korea}
\affiliation{Department of Physics, Xiamen University, Xiamen 361005, Fujian Province, China}

\date{\today}

\begin{abstract}
  Non-Hermitian dynamics possesses distinct left and right eigenstates, but their normalization freedom obscures a direct probabilistic interpretation.
  Starting from a globally Hermitian system governed by the standard Born rule, we generalize dynamical quantum phase transitions (DQPTs) from a reference state to a physical subspace.
  This generalization provides a clear probabilistic interpretation of left and right states: when each state is normalized to itself, their mutual overlap determines a directly accessible dynamical transition rate.
  We discuss the new phenomena governed by this emergent non-biorthogonal Born rule and further identify a class of exceptional-point DQPTs (EP-DQPTs), at which the eigenstates of the critical propagator coalesce.
  Our theory places the probabilistic interpretation of non-Hermitian quantum mechanics on the firm footing of Hermitian quantum mechanics.
\end{abstract}

\maketitle

{\em Introduction.}---The standard Born rule gives a probabilistic interpretation of quantum mechanics through the overlaps between different orthonormal states~\cite{Born1926,Gleason1957}.
Non-Hermitian effective Hamiltonians, ubiquitous in open quantum systems~\cite{Rotter2009,Ashida2020} and correlated matter~\cite{Kozii2017,Shen2018,Yoshida2020,MichishitaPeters2020}, break this structure: they possess non-orthonormal left and right eigenstates~\cite{Brody2014,Bergholtz2021},
\begin{equation}
  H_{\mathrm{NH}}|R_n\rangle=E_n|R_n\rangle,\qquad
  \langle L_n|H_{\mathrm{NH}}=E_n\langle L_n|.
  \label{eq:biortho_eigen}
\end{equation}
Although both families are widely used in non-Hermitian physics, their probabilistic interpretation has remained unsettled~\cite{Scholtz1992,Brody2014,Mostafazadeh2010,Wang2026GaugeInvariantNonHermitianDPT,Fu2026ProbabilityBalance}.

Concretely, one popular convention is the biorthogonal normalization~\cite{Moiseyev2011,Brody2014,Mostafazadeh2002}:
\begin{equation}
  \langle L_m|R_n\rangle=\delta_{mn}.
  \label{eq:biortho_norm}
\end{equation}
One readily checks that this relation is preserved by the rescaling $|R_n\rangle\to c_n|R_n\rangle$, $\langle L_n|\to c_n^{-1}\langle L_n|$, which changes every norm $\langle R_n|R_n\rangle$ and $\langle L_n|L_n\rangle$~\cite{Brody2014,Edvardsson2022}.
This gauge freedom obstructs any direct probability interpretation: the one-sided weight $|\langle L_n|\psi\rangle|^2$ is not gauge invariant; the bilinear $\langle\psi|R_n\rangle\langle L_n|\psi\rangle$ is invariant but generically complex.
Biorthogonal quantum mechanics introduces an associated left state to obtain real, nonnegative probabilities~\cite{Brody2014}.
This construction assigns zero transition probability between distinct right eigenstates.
In the standard measurement framework advocated in Ref.~\cite{Meden2024Methodological}, however, preparing one normalized right eigenstate and projecting onto another gives $|\langle R_m|R_n\rangle|^2$, which is generally nonzero.
These different answers reflect different choices of inner product and measurement, whose physical meaning cannot be determined by the non-Hermitian eigenproblem alone.

To make progress, it is natural to start where probabilities are unambiguous: a globally Hermitian system~\cite{Ashida2020,Wu2019}.
This line of thought has already provided some useful insights: when a closed system is partitioned into a subsystem $S$ and an environment $E$, the wavefunction component within $S$ exactly obeys a time-dependent non-Hermitian Schr\"{o}dinger equation with $\Heff(t) = H_S + \Sigma_{\rm inst}(t)$~\cite{Rotter2009,Ashida2020,YangNHQMI}.
As reviewed in Appendix~\ref{app:self_energy_no_go}, the instantaneous self-energy $\Sigma_{\rm inst}(t)$ encodes the back-action of the environment~\cite{Feshbach1958,Rotter2009}.
Every probability is then fixed by the standard Born rule of the parent Hilbert space~\cite{Born1926,Gleason1957}, which provides a rigorous foundation for the probabilistic interpretation of non-Hermitian dynamics.

A key question nevertheless remains: to date, typical non-Hermitian experiments only prepare and measure states in orthogonal bases, with the biorthogonal basis serving mainly as a theoretical tool that experiments can hardly detect directly~\cite{Xiao2020,Wu2019,WangMa2025,Schomerus2020Nonreciprocal,PhysRevLett.125.186802,Siegman1989,Kunst2018,Wiersig2023,Kullig2025GeneralizedPetermann,Bid2025UniformResponse}.
Therefore, it is natural to ask whether there are some physical processes that can dynamically select the left and right eigenstates and can help us to reveal their roles in a Born interpretation.

In this Letter, we show that dynamical quantum phase transitions (DQPTs) provide precisely such a quantity, once suitably generalized.
DQPTs have been studied across a wide range of settings, including sudden quenches~\cite{Heyl2013,Heyl2018,Zhou2021FiniteSizeZeros,Zeng2023DynamicalSingularity}, slow ramps~\cite{Zhang2024UniversalScalingDQPT,Zhang2026FateDQPT}, Floquet driving~\cite{Yang2019FloquetDQPT,Zhou2021FloquetDQPTQuenched,Naji2022EngineeringFloquetDQPT}, topological and disordered systems~\cite{Wang2018DQPTSteadyState,Khan2023CorrelationDQPT,Khan2024LinearScaleQuench}, and non-Hermitian extensions~\cite{Qiu2019PTSymmetricQuench,Zhou2018NonHermitianLatticeDQPT,Zhou2021NonHermitianTopologyDQPT,Fu2025AnatomyNonHermitianDQPT,Jing2024BiorthogonalDQPT}, with experimental realizations in trapped ions, cold atoms, and photonic systems~\cite{Jurcevic2017,Flaschner2018,Guo2019DQPTQubit,Wang2019PhotonicDQPT}.
The standard diagnostic is the vanishing of the Loschmidt echo amplitude,
\begin{equation}
  \mathcal G_{\psi}(t)=\langle\psi|U(t)|\psi\rangle,\qquad
  \mathcal L_{\psi}(t) = |\mathcal G_{\psi}(t)|^2,
  \label{eq:loschmidt_echo}
\end{equation}
where $\mathcal L_{\psi}(t)$ is the return probability of an evolved reference state.
A vanishing echo means that the evolved state is orthogonal to its initial state; quantum speed limits set a lower bound on the time required~\cite{Liu2021BhatiaDavisQSL}.
For a many-body system of size $N$, the return rate is the negative logarithm of the echo per degree of freedom, $r_N(t)=-N^{-1}\ln\mathcal L_\psi(t)$.
In the thermodynamic limit, this rate can develop nonanalytic features, such as cusps, which signal a DQPT~\cite{Heyl2013}.
As a diagnostic, however, this criterion has two limitations: it depends on the chosen $|\psi\rangle$~\cite{Andraschko2014,Xu2024QuantumCoherenceDQPT}, and a vanishing echo reveals nothing about transitions between different states at $t_c$.

We will show that both limitations mentioned above can be removed by promoting the reference state to a physical subspace $S$ (a set of lattice sites or a spin sector) with an orthonormal basis $\{|m\rangle\}_{m=1}^{d}$, $d=\dim S$.
In terms of the subspace propagator, a \emph{subspace DQPT} occurs when
\begin{equation}
  [\US(t)]_{mn} = \langle m|\ee^{-\ii Ht}|n\rangle, \qquad \det\US(t_c) = 0,
  \label{eq:subspace_dqpt_intro}
\end{equation}
a criterion that is basis independent within $S$, reduces to the ordinary Loschmidt zero for $d = 1$, and guarantees that at least one state in $S$ has vanishing echo at $t_c$.
In Appendix~\ref{app:self_energy_no_go}, we will show that when $\det\US(t_c)=0$, $\Sigma_{\rm inst}(t)$ (and hence $\Heff(t)$) diverges as $t\to t_c$, identifying that the DQPT is exactly a singularity of the projected dynamics.

The matrix structure of $\Uc \equiv \US(t_c)$ raises two questions the scalar echo cannot pose: how many states in $S$ lose their echo at $t_c$, and what governs the transition probabilities among the rest?
We show that both answers are fixed by the left and right eigenstates of $\Uc$.
This delivers a direct physical interpretation of the states: when each eigenstate is separately unit-normalized in the physical norm, i.e., $\langle L_1|L_1\rangle=\langle R_1|R_1\rangle = 1$, the overlap $|\langle L_1|R_1\rangle|$, which we term the dynamical phase rigidity~\cite{vanLangen1997,Wiersig2023,Schomerus2024}, directly governs a physically accessible transition probability that can be detected in experiments.
We therefore call this a non-biorthogonal, rather than biorthogonal, Born rule~\cite{nonbiorthonormal_terminology_SM}.

\begin{figure}[t!]
  \centering
  \includegraphics[width=\columnwidth]{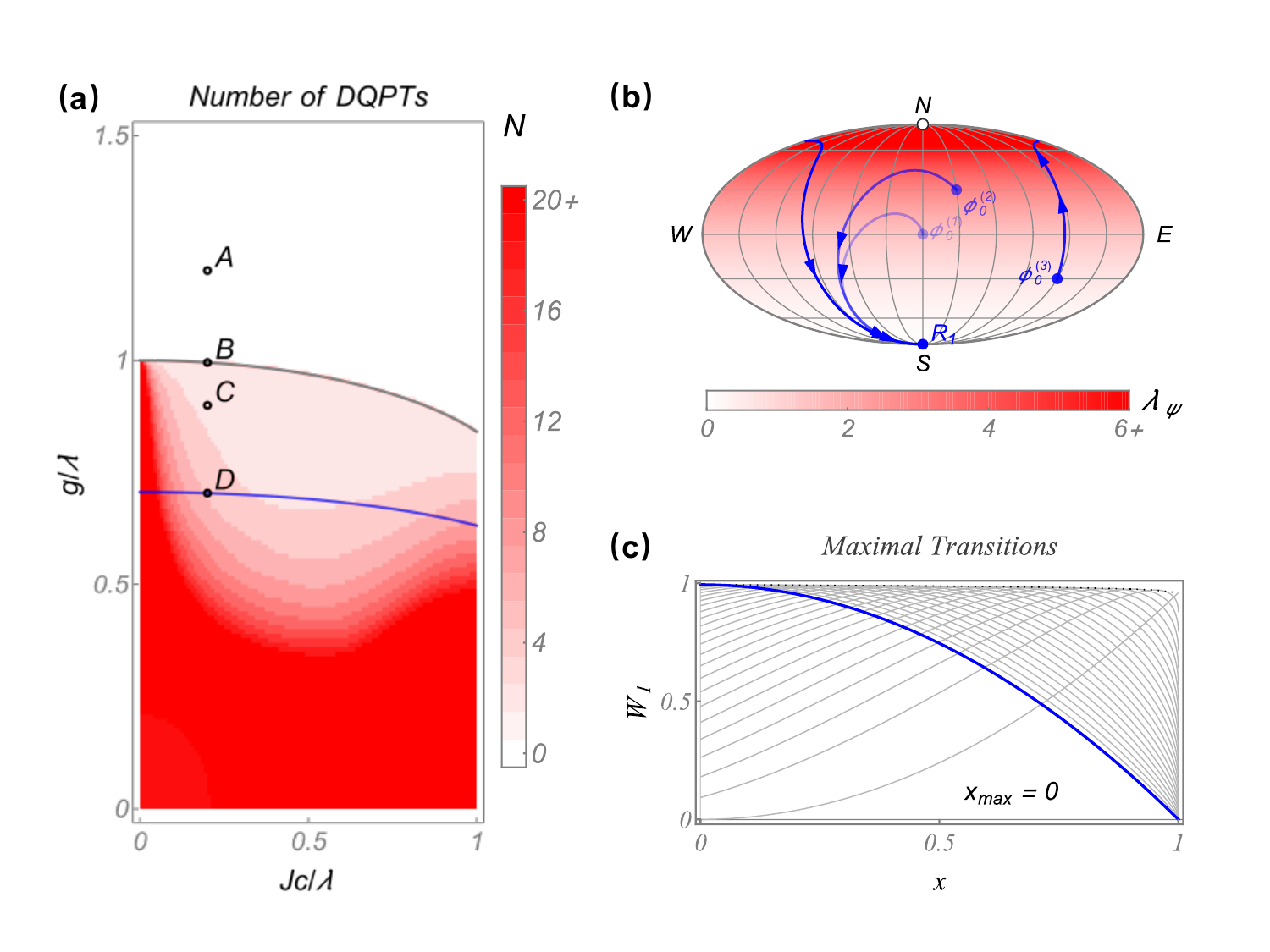}
  \caption{\label{fig:dqpt_phase_dynamics}
    Phase diagram and the orthogonal DQPT.
    (a) Number $N_{\mathrm{DQPT}}$ of zeros of $\det\US(t)$ within the observation window $\lambda t \le 60$, versus $J_c/\lambda$ and $g/\lambda$.
    The gray and blue curves mark the DQPT boundary and the EP-DQPT locus; points A--D at $J_c/\lambda=0.2$ mark the no-DQPT ($g/\lambda=1.2$), boundary ($0.9953$), generic ($0.9$), and EP-DQPT ($0.7047$) parameter sets.
    (b) Critical-time Loschmidt-echo landscape $\lambda_\psi=-\ln\mathcal L_\psi(t_c)$ on the Bloch sphere (Mollweide projection) at the boundary point B.
    The unique zero-echo state sits at the north pole, and the projected trajectories of distinct initial states $\phi_0^{(1,2,3)}$ all funnel into the output $|R_1\rangle$ at the south pole.
    (c) Transition probability $W_1=|\langle R_1|\Uc|\psi\rangle|^2$ versus $x=|\langle R_1|\psi\rangle|$ at the same point (blue), maximal at $x_{\max}=1$.
    Thin gray curves show the same transition probability as $g/\lambda$ decreases from the boundary toward the EP-DQPT point: the peak drifts continuously from $x=1$ to $x=0$ [cf.~Figs.~\ref{fig:nonorthogonal_ep_diagnostics}(c) and \ref{fig:nonorthogonal_ep_diagnostics}(d)].}
\end{figure}

Beyond this rule, two further structures emerge.
Deeper in the DQPT phase, the eigenstates of $\Uc$ coalesce along a critical locus, defining an EP-DQPT that extends the notion of exceptional points (EPs)~\cite{Heiss2012,Miri2019,Bergholtz2021,Fu2024BlochSphere,Wang2023NonlinearLZSM} to DQPTs.
Secondly, within the parameter range of Fig.~\ref{fig:dqpt_phase_dynamics}(a), the displayed boundary component belongs to the orthogonal class.

  {\em Model.}---To develop these insights in a concrete setting, we consider a two-level subsystem $S = \{|A\rangle, |B\rangle\}$ coupled to a one-dimensional tight-binding chain that serves as the environment.
The total Hamiltonian $H = H_S + H_E + H_{\mathrm{S\text{-}E}}$ is globally Hermitian.
We focus first on the single-particle (first-quantized) setup, which can also be realized experimentally in various classical wave platforms where the probabilistic interpretation transfers to intensity measurements~\cite{ElGanainy2018NatPhys,Ding2022NatRevPhys,single_particle_scope_SM}.
Here $H_S = g\bigl(|A\rangle\langle B| + \mathrm{H.c.}\bigr)$ couples the two levels coherently, $H_E = -J_c\sum_{n=1}^{L}(|n\rangle\langle n{+}1| + \mathrm{H.c.})$ is a tight-binding chain of $L$ sites with bandwidth $4J_c$, and $H_{\mathrm{S\text{-}E}} = \lambda(|B\rangle\langle 1| + \mathrm{H.c.})$ attaches $|B\rangle$ to the chain's boundary site $|1\rangle$.
The dimensionless control parameters are $g/\lambda$ and $J_c/\lambda$.

After the environment degrees of freedom are eliminated in the thermodynamic limit ($L\to\infty$), the subspace propagator is obtained via the resolvent~\cite{YangNHQMI}:
\begin{equation}
  \US(t) = \oint \frac{\mathrm{d}z}{2\pi\ii}\,\ee^{-\ii zt} \left[z-\begin{pmatrix} 0 & g \\[1ex] g & \lambda^2 G_E^R(z) \end{pmatrix}\right]^{-1},
  \label{eq:USS_resolvent}
\end{equation}
where $G_E^R(z) = (z - \sqrt{z^2 - 4J_c^2})/(2J_c^2)$ is the exact boundary Green's function of the semi-infinite chain {\cite{Datta1995}}.
We establish a no-go theorem in Appendix~\ref{app:self_energy_no_go}: a subspace DQPT requires the instantaneous self-energy $\Sigma_{\rm inst}(t)$ to be strongly time dependent and is therefore impossible in any Markovian subsystem.
These non-Markovian features are most pronounced in the narrow-band regime ($J_c < \lambda$), on which we focus in the following discussion~\cite{narrow_band_memory_SM}.

Fig.~\ref{fig:dqpt_phase_dynamics} (a) maps $N_{\mathrm{DQPT}}$, the number of zeros of $\det\US(t)$ within the fixed observation window $0\leq\lambda t\leq60$, as a function of parameters $g$ and $J_c$; a nonzero count identifies the DQPT phase.
The gray and blue curves mark the DQPT boundary and the EP-DQPT locus.
Throughout, we fix $J_c/\lambda=0.2$, and follow the four representative points marked A--D in Fig.~\ref{fig:dqpt_phase_dynamics} (a): no DQPT ($g/\lambda=1.2$), boundary ($g/\lambda=0.9953$), generic ($g/\lambda=0.9$), and EP-DQPT ($g/\lambda=0.7047$), for which $|\det U_{SS}(t)|$ is shown in Appendix~\ref{app:det_dynamics}.

{\em Orthogonal DQPT.}---We first examine the boundary point B, whose determinant touches zero only once at $t_c \approx 2.238/\lambda$.
Numerically, $\Uc$ is found to be normal at this point, so its normalized eigenstates form an orthonormal basis of $S$.
Write $|R_1\rangle$ and $|R_2\rangle$ for the eigenstates with nonzero and zero eigenvalue; the spectral representation then reads
\begin{equation}
  \Uc = \eta\,|R_1\rangle\langle R_1| + 0\cdot|R_2\rangle\langle R_2|.
  \label{eq:Uc_spectral_boundary}
\end{equation}

Three signatures follow from the standard Born rule.
(i)~The Loschmidt amplitude $\langle\psi|\Uc|\psi\rangle=\eta|\langle R_1|\psi\rangle|^2$ vanishes if and only if $|\psi\rangle=|R_2\rangle$: exactly one zero-echo state appears on the Bloch sphere, consistent with $\det\Uc=0$ [north pole of Fig.~\ref{fig:dqpt_phase_dynamics}(b)].
Beyond this consistency check, the matrix structure of $\Uc$ yields two features inaccessible to the single-state formulation.
(ii)~Every non-kernel input obeys $\Uc|\psi\rangle=\eta\langle R_1|\psi\rangle|R_1\rangle\propto|R_1\rangle$: normalized trajectories funnel into a single output ray, independently of the initial direction [trajectories in Fig.~\ref{fig:dqpt_phase_dynamics}(b)].
(iii)~As a function of the overlap $x\equiv|\langle R_1|\psi\rangle|$, the transition probability $W(R_1\!\leftarrow\!\psi)=|\langle R_1|\Uc|\psi\rangle|^2=|\eta|^2|\langle R_1|\psi\rangle|^2$ is maximal at $x=1$: the most efficient input is the selected output itself [Fig.~\ref{fig:dqpt_phase_dynamics}(c)].

Because the eigenframe is strictly orthogonal, we call this an \emph{orthogonal DQPT}; it involves no departure from Hermitian intuition.
The physical origin of this orthogonal DQPT at the phase boundary is discussed in the Supplemental Material~\cite{orthogonal_boundary_SM}.
Furthermore, if $\eta=0$, then $\Uc=0$, which is a \emph{degenerate DQPT}, and every state loses its echo.

\begin{figure}[t!]
  \centering
  \includegraphics[width=\columnwidth]{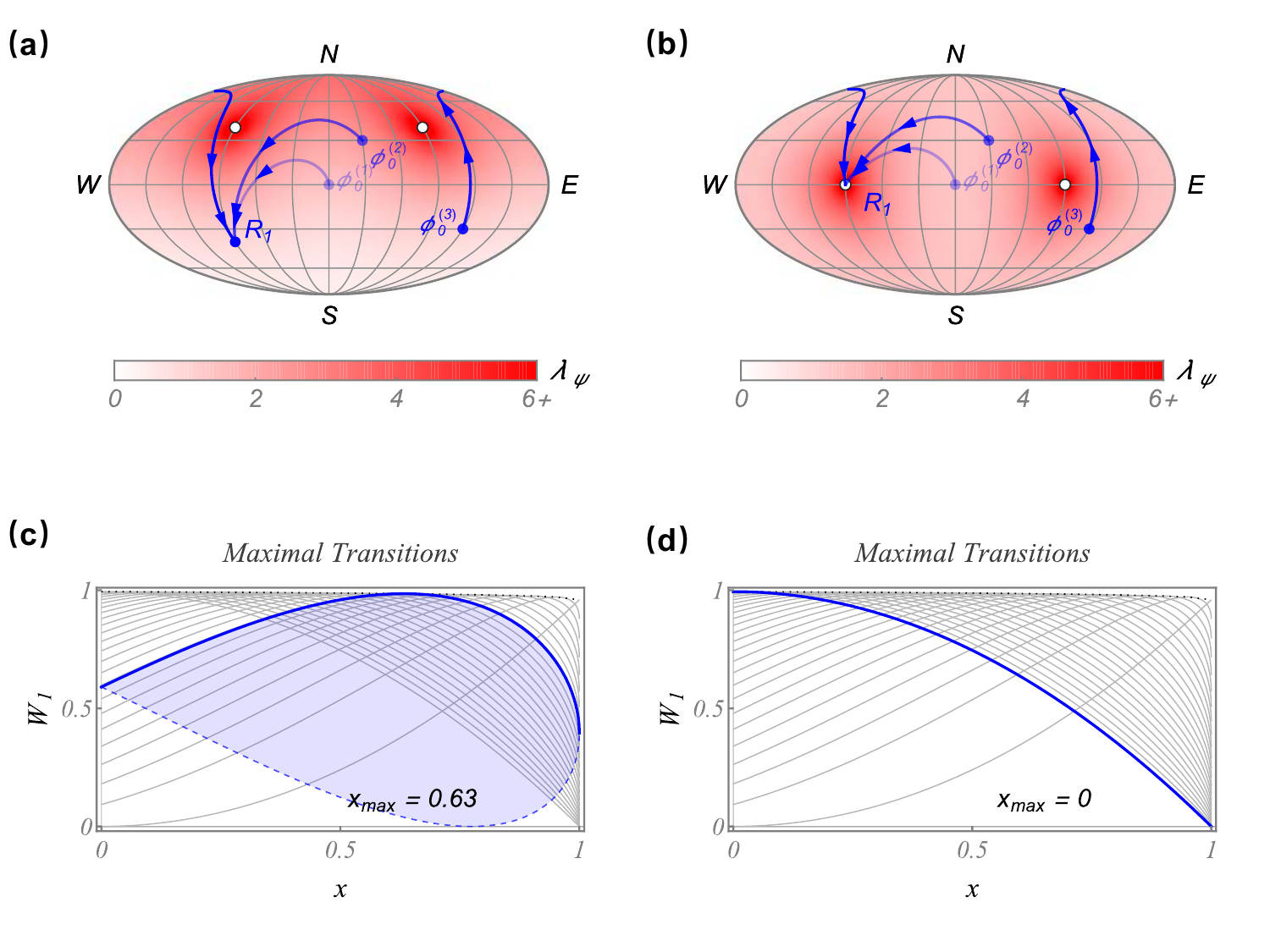}
  \caption{\label{fig:nonorthogonal_ep_diagnostics}
    Critical-time Loschmidt-echo landscapes on the Bloch sphere [(a),(b)] and transition probability $W_1$ versus $x$ [(c),(d)], for the generic non-biorthogonal DQPT point [(a),(c)] and the EP-DQPT point [(b),(d)].}
\end{figure}

{\em Non-biorthogonal DQPT.}---We now move to the generic interior point C and repeat the same three diagnostics.
(i)~Two distinct states on the Bloch sphere now lose their echo [white markers in Fig.~\ref{fig:nonorthogonal_ep_diagnostics}(a)], while generic states retain finite echo.
(ii)~The funnel nevertheless persists: distinct initial states still converge to a single output ray [trajectories in Fig.~\ref{fig:nonorthogonal_ep_diagnostics}(a)].
(iii)~The maximum of $W(R_1\!\leftarrow\!\psi)$ shifts away from $x=1$ to $x_{\max}=0.63$ [Fig.~\ref{fig:nonorthogonal_ep_diagnostics}(c)]: the most efficient input is no longer the output itself.
Within the orthogonal framework of Eq.~\eqref{eq:Uc_spectral_boundary}, these observations are mutually contradictory: two independent zero-echo states would force $\Uc=0$, yet the surviving funnel and the finite generic echoes show $\Uc\neq0$.

The contradiction is resolved by a single structural change: $\Uc$ remains rank one but becomes non-normal.
With the output $|R_1\rangle$ and the receptive input $|L_1\rangle$, which are normalized to themselves, i.e., $\langle R_1|R_1\rangle = \langle L_1|L_1\rangle = 1$, $U_c$ can be written as
\begin{equation}
  \Uc = \kappa\,|R_1\rangle\langle L_1|+ 0\cdot|R_2\rangle\langle L_2|, \qquad 0<|\kappa|\leq 1,
  \label{eq:Uc_spectral}
\end{equation}
where $|\kappa|$ is the unique nonzero singular value and the nonzero eigenvalue is $\eta=\kappa\langle L_1|R_1\rangle$ (for a detailed discussion, please refer to Appendix~\ref{app:normalization}).
We emphasize that since $H_{\rm eff}(t)$ is time-dependent, $\{|R_i\rangle,|L_i\rangle\}$ are generally not the eigenstates of $H_{\rm eff}(t)$~\cite{YangNHQMI}.
The non-biorthogonality probed at the DQPT is thus a property of the full time-integrated dynamics, not of any single-time generator.
The amplitude $\langle\psi|\Uc|\psi\rangle=\kappa\langle\psi|R_1\rangle\langle L_1|\psi\rangle$ now vanishes for two distinct states, the right null state $|R_2\rangle\perp|L_1\rangle$ and the left null state $|L_2\rangle\perp|R_1\rangle$, which explains~(i).
The funnel~(ii) persists because $\Uc|\psi\rangle=\kappa\langle L_1|\psi\rangle|R_1\rangle$ for every non-kernel input, and the shift~(iii) follows from
\begin{equation}
  W(R_1\!\leftarrow\!\psi) = |\kappa|^2\,|\langle L_1|\psi\rangle|^2,
  \label{eq:biortho_born}
\end{equation}
whose maximum sits at $|\psi\rangle=|L_1\rangle$, i.e., at $x_{\max}=|\langle L_1|R_1\rangle|$ on the plotted axis.

The peak position $x_{\max}=|\langle L_1|R_1\rangle|$ thus directly quantifies the left--right overlap, and we call Eq.~\eqref{eq:biortho_born} the \emph{emergent non-biorthogonal Born rule}.
It is not an alternative probability postulate; the global Born rule of the parent Hermitian system remains intact.
What changes at the subspace DQPT is the geometry: the receptive input differs from the selected output.
This answers the question posed in the introduction: the non-biorthogonal frame is not imposed by preparation but dynamically selected by the critical propagator, and its non-biorthogonality is carried into a measurable probability.

\begin{figure}[t!]
  \centering
  \includegraphics[width=1.\columnwidth]{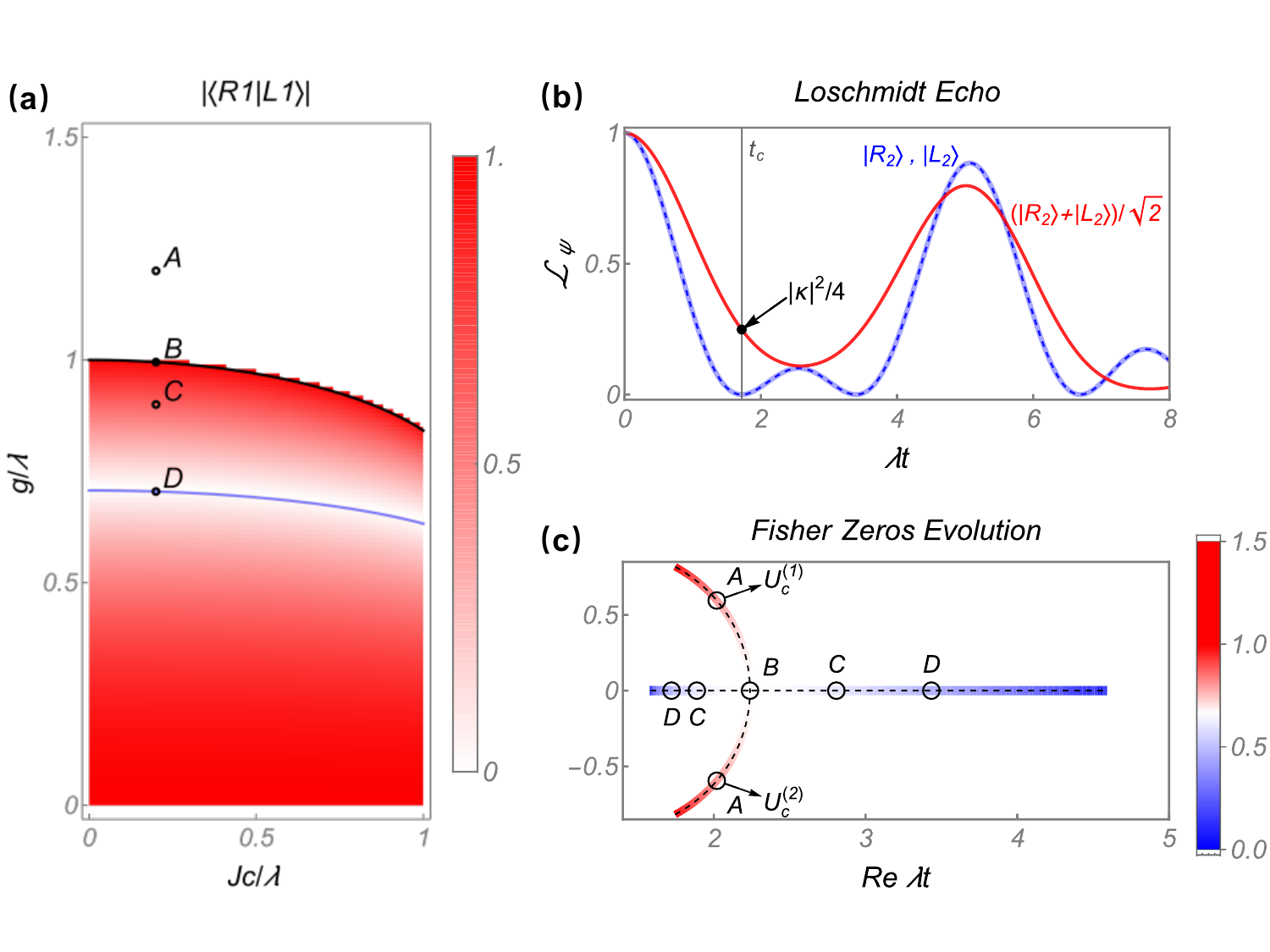}
  \caption{\label{fig:r1r2_overlap_phase}
    Critical-state geometry and critical-time structure.
    (a)~Overlap $|\langle L_1|R_1\rangle|$ across the phase diagram; curves and points A--D as in Fig.~\ref{fig:dqpt_phase_dynamics}(a).
    From the DQPT boundary to the EP-DQPT locus it falls from unity to zero, obeying $|\langle R_1|R_2\rangle|^2+|\langle L_1|R_1\rangle|^2=1$.
    (b)~Loschmidt echoes at the EP point D: those of $|R_2\rangle$ and $|L_2\rangle$ coincide and vanish at $t_c$, whereas their equal superposition has $\mathcal L(t_c)=|\kappa|^2/4$.
    (c)~Fisher zeros of $\det\US$ in the complex-$\lambda t$ plane as $g/\lambda$ varies at $J_c/\lambda=0.2$; colors encode $g/\lambda$, dashed curves trace the zero trajectories, and circles mark points A--D.}
\end{figure}

  {\em Exceptional-point DQPT.}---The input--output misalignment is largest along the blue curve in Fig.~\ref{fig:dqpt_phase_dynamics}(a), where the receptive input is orthogonal to the selected output, $\langle L_1|R_1\rangle=0$ [Fig.~\ref{fig:r1r2_overlap_phase}(a)].
The critical propagator then squares to zero without vanishing itself:
\begin{equation}
  \Uc^2 = \kappa^2\,|R_1\rangle\langle L_1|R_1\rangle\langle L_1| = 0, \qquad \Uc \neq 0.
  \label{eq:EP2_propagator}
\end{equation}
Such a $\Uc$ cannot be diagonalized: its two eigenstates coalesce, $|R_2\rangle\propto|R_1\rangle$ and $|L_2\rangle\propto|L_1\rangle$, so $\Uc$ sits at an EP.
We call a DQPT on this curve an \emph{EP-DQPT}.

Here the departure from Hermitian intuition becomes maximal.
Although every state with a nonzero critical-time output is funneled onto $|R_1\rangle$, an initial state prepared in $|R_1\rangle$ itself has zero echo at $t_c$, $\mathcal L_{R_1}(t_c)=0$ [in Fig.~\ref{fig:nonorthogonal_ep_diagnostics}(b), $|R_1\rangle$ sits directly on a zero-echo marker].
To reach $|R_1\rangle$ most efficiently, one must start orthogonal to it: the peak of $W(R_1\!\leftarrow\!\psi)$ sits at $x_{\max}=0$ [Fig.~\ref{fig:nonorthogonal_ep_diagnostics}(d)].
%Nothing here modifies quantum mechanics: both statements follow from the standard Born rule of the parent Hermitian system.
%It is the rank-one projection that inverts Hermitian intuition.

Unlike conventional EPs of non-Hermitian Hamiltonians, the EP-DQPT is a geometric crossover within the DQPT phase.
The EP-DQPT locus does not by itself define a boundary between regions with different numbers of critical times.
On this locus, $\Uc$ remains rank one but exhibits a striking configuration: two mutually orthogonal states, $|R_1\rangle$ and $|L_1\rangle$, that both carry zero echo.
These exceptional properties are basis independent: under a time-independent unitary change of basis, $\Uc$ transforms by similarity and retains its defectiveness and Jordan structure. Under a time-dependent gauge transformation, the same physical EP is recovered after identifying the final frame with the initial one~\cite{time_dependent_gauge_SM}.

{\em Non-biorthogonal revival.}---At the EP-DQPT, the orthogonal states $|R_1\rangle$ and $|L_1\rangle$ both have zero echo at $t_c$, yet their superposition does not.
For $|\psi\rangle=\alpha|R_1\rangle+\beta|L_1\rangle$ with $|\alpha|^2+|\beta|^2=1$, the propagator annihilates the $|R_1\rangle$ component and sends the $|L_1\rangle$ component to $|R_1\rangle$, so the return amplitude is
\begin{equation}
  \langle\psi|\Uc|\psi\rangle=\kappa\,\langle\psi|R_1\rangle\langle L_1|\psi\rangle=\kappa\,\alpha^*\beta.
  \label{eq:biortho_revival}
\end{equation}
The echo $\mathcal L_\psi(t_c)=|\kappa|^2|\alpha|^2|\beta|^2$ is finite whenever both components are present and reaches its maximum $|\kappa|^2/4$ at equal weights, $|\alpha|=|\beta|=1/\sqrt{2}$.
We call this \emph{non-biorthogonal revival}: superposing two zero-echo states restores a finite echo.
Figure~\ref{fig:r1r2_overlap_phase}(b) shows this effect at the EP point D.

  %This result also makes the exceptional structure of $\Uc$ transparent.
  %Because $|\kappa|$ is its unique nonzero singular value, $\Uc$ remains nonzero at the EP.
  %However, the EP condition $\langle L_1|R_1\rangle=0$ forces its only potentially nonzero eigenvalue, $\eta=\kappa\langle L_1|R_1\rangle$, to vanish.
  %Thus both eigenvalues are zero, while $\Uc\neq0$ and $\Uc^2=0$ [Eq.~\eqref{eq:EP2_propagator}]; $\Uc$ is a nonzero rank-one nilpotent operator.
  %the effect at the EP point D: the echoes of the two zero-echo states vanish at $t_c$, while that of their equal superposition stays finite.
  %Away from the EP the amplitude in Eq.~\ref{eq:biortho_revival} acquires an additional factor $\langle R_2|R_1\rangle\langle L_1|L_2\rangle$, an off-diagonal overlap of the form known from non-Hermitian eigenstate geometry~\cite{ChalkerMehlig1998,Schomerus2024}, so the revival persists throughout the non-biorthogonal phase.
  %Experimentally, at any fixed DQPT point, $|\kappa|$ can be extracted from the maximal output probability over all initial states, $\max_{\psi}W(R_1\!\leftarrow\!\psi)=|\kappa|^2$.

  {\em Adjoint twins.}---We finally discuss the Fisher zeros, i.e., the zeros of $\det\US(t)$ continued to complex time $t$~\cite{Heyl2013} (see also Appendix~\ref{app:thermodynamic_analogy}), as shown in Fig.~\ref{fig:r1r2_overlap_phase}(c).
It can be proved that as long as the total Hamiltonian has chiral symmetry, $\det\US(t)$ must be real.
For the parameters chosen in Fig.~\ref{fig:dqpt_phase_dynamics} (a), as $g$ varies, the Fisher zeros undergo a ``PT''-like transition in the complex-$t$ plane [Fig.~\ref{fig:r1r2_overlap_phase}(c)]: a complex-conjugate pair (point A) meets the real axis at the phase boundary (point B) and splits into pairs of real critical times (points C and D).
These critical times are zeros of $\det\US$, not eigenvalues of $\US$, so the collision is not an EP of $\US$.
Numerically, we found that within the plotted parameter range, the critical propagators at the two members $t_{c1}$ and $t_{c2}$ of each emerging pair form \emph{adjoint twins},
\begin{equation}
  \Uc^{(2)} \simeq \bigl(\Uc^{(1)}\bigr)^{\dagger}.
  \label{eq:adjoint_twins}
\end{equation}
At the phase boundary, the twins merge at the same critical time, which implies that $\Uc=\Uc^{\dagger}$, indicating an orthogonal DQPT on the boundary.
Furthermore, this merger also leads to $\det\US\sim(t-t_c)^2$, giving $\alpha=2$, whereas interior zeros are simple crossings with $\alpha=1$.

\begin{table}[t]
  \caption{\label{tab:classification}Fourfold classification of subspace DQPTs for $\dim S = 2$.
    $N_z$: number of zero-echo states; $x_{\rm max}$: overlap at which transition probability is maximal; $\alpha$: critical exponent of $|\det\US(t)| \sim |t-t_c|^\alpha$.}
  \begin{ruledtabular}
    \footnotesize
    \setlength{\tabcolsep}{3pt}
    \begin{tabular}{lccccc}
      Class            & $|\langle L_1|R_1\rangle|$ & $N_z$    & Revival      & $x_{\rm max}$              & $\alpha$ \\
      \hline
      Degenerate       & --                         & $\infty$ & --           & --                         & 2        \\
      Orthogonal       & $1$                        & 1        & $\times$     & $1$                        & 2        \\
      Non-biorthogonal & $(0,1)$                    & 2        & $\checkmark$ & $|\langle L_1|R_1\rangle|$ & 1        \\
      Exceptional      & $0$                        & 2        & $\checkmark$ & $0$                        & 1        \\
    \end{tabular}
  \end{ruledtabular}
\end{table}

{\em Summary and discussion.}---In summary, we identified a non-biorthogonal Born rule that gives precise meaning to the role of right and left eigenstates in terms of subspace DQPTs. Tab.~\ref{tab:classification} summarizes the concrete manifestations of this rule in different regimes, amounting to a fourfold classification of subspace DQPTs with $\dim S=2$.
This subspace DQPT, defined by the criterion $\det\US(t_c)=0$, is invariant under both time-dependent basis choices at initial and final times, as well as under subsystem similarity transformations, and requires non-Markovian memory: the instantaneous self-energy must diverge at the critical time.
Every $2\times2$ critical subspace propagator belongs to exactly one of these classes in Tab.~\ref{tab:classification}.
The second column gives the structural classification criterion.
The remaining four columns, $N_z$, revival, $x_{\max}$, and $\alpha$, provide experimentally accessible diagnostics for identifying the class in terms of the non-biorthogonal Born rule.

Several generalizations follow naturally.
First, for an $m$-level subspace, $k$ simultaneously vanishing eigenvalues define a $k$th-order DQPT with up to $2k$ zero-echo states~\cite{higher_order_dqpt_SM}.
Second, even for general quantum many-body systems, the single-particle sector itself forms a natural subspace.
Here, the retarded Green's function obeys the same Dyson structure as Eq.~\eqref{eq:USS_resolvent}, with the self-energy now supplied by interactions.
The criterion
\begin{equation}
  \det G^R(t_c)=0
\end{equation}
then defines an interaction-induced \emph{quasiparticle DQPT}.
Wherever the self-energy is strongly frequency dependent, as in Kondo impurities, Mott insulators, and heavy-fermion compounds, such zeros may occur generically, extending the non-biorthogonal Born rule to correlated matter.

\bibliography{references}

\onecolumngrid
\begin{center}
  \bfseries End Matter
\end{center}
\twocolumngrid

\refstepcounter{appref}\label{app:self_energy_no_go}
{\em Appendix A: Instantaneous self-energy and the no-go theorem.}---Whenever $\US(t)$ is invertible, the projected dynamics admits an exact time-local generator~\cite{YangNHQMI}:
\begin{equation}
  \Heff(t)=\ii\bigl[\partial_t\US(t)\bigr]\US^{-1}(t)=H_S+\Sigma_{\rm inst}(t),
  \label{eq:heff_def}
\end{equation}
where the instantaneous self-energy $\Sigma_{\rm inst}(t)$ encodes the environmental back-action.
For $t>0$, its exact frequency-domain representation is~\cite{YangNHQMI}
\begin{equation}
  \begin{aligned}
    \Sigma_{\rm inst}(t)
     & =\left[\frac{\ii}{2\pi}\int_{-\infty}^{\infty}\!\mathrm{d}\omega\,
    \ee^{-\ii\omega t}\Sigma_S^R(\omega)G_S^R(\omega)\right]                        \\
     & \quad\times\left[\frac{\ii}{2\pi}\int_{-\infty}^{\infty}\!\mathrm{d}\omega\,
      \ee^{-\ii\omega t}G_S^R(\omega)\right]^{-1},
  \end{aligned}
  \label{eq:sigma_inst_frequency}
\end{equation}
where $G_S^R(\omega)=[(\omega+\ii0^+)I_S-H_S-\Sigma_S^R(\omega)]^{-1}$ and $\Sigma_S^R(\omega)=H_{SE}[(\omega+\ii0^+)I_E-H_E]^{-1}H_{ES}$, with $H_{XY}=P_XHP_Y$ and $P_X$ the projector onto $X\in\{S,E\}$.
Taking the trace of Eq.~\eqref{eq:heff_def} gives $\Tr\Heff(t)=\ii\,\partial_t\ln\det\US(t)$, i.e., Jacobi's formula:
\begin{equation}
  \det\US(t)=\exp\Bigl[-\ii\int_0^t\Tr\Heff(t')\,\mathrm{d}t'\Bigr].
  \label{eq:jacobi}
\end{equation}
Because $H_S$ is Hermitian, only $\mathrm{Im}\Tr\Sigma_{\rm inst}$ contributes to the modulus.
A subspace DQPT therefore announces itself in the self-energy: as $t\to t_c$, $\ln\det\US\to-\infty$, so $\Sigma_{\rm inst}(t)$ and $H_{\rm eff}(t)$ must diverge.
The no-go theorem follows as the contrapositive.
For a closed system, $\det U(t)=\ee^{-\ii t\Tr H}$ has unit modulus and never vanishes; echo zeros on the full space, such as $\langle0|\sigma_x|0\rangle=0$, reflect orthogonality rather than rank loss (Appendix~\ref{app:higher_order_dqpt}).
If the subsystem dynamics is generated by any bounded time-local generator, such as a phenomenological non-Hermitian Hamiltonian or a Markovian master equation, then $|\det\US(t)|$ remains nonzero at every finite time.
In the wide-band limit, the instantaneous self-energy is the constant $\Sigma_{\rm inst}=-\ii\Gamma|B\rangle\langle B|$.
One finds that such a constant instantaneous self-energy gives $\det\US(t)=\ee^{-\Gamma t}$.
Conversely, an observed zero of $\det\US$ falsifies every bounded time-local non-Hermitian description of the subsystem and witnesses non-Markovian memory backflow.

\begin{figure}[t!]
  \centering
  \includegraphics[width=\columnwidth]{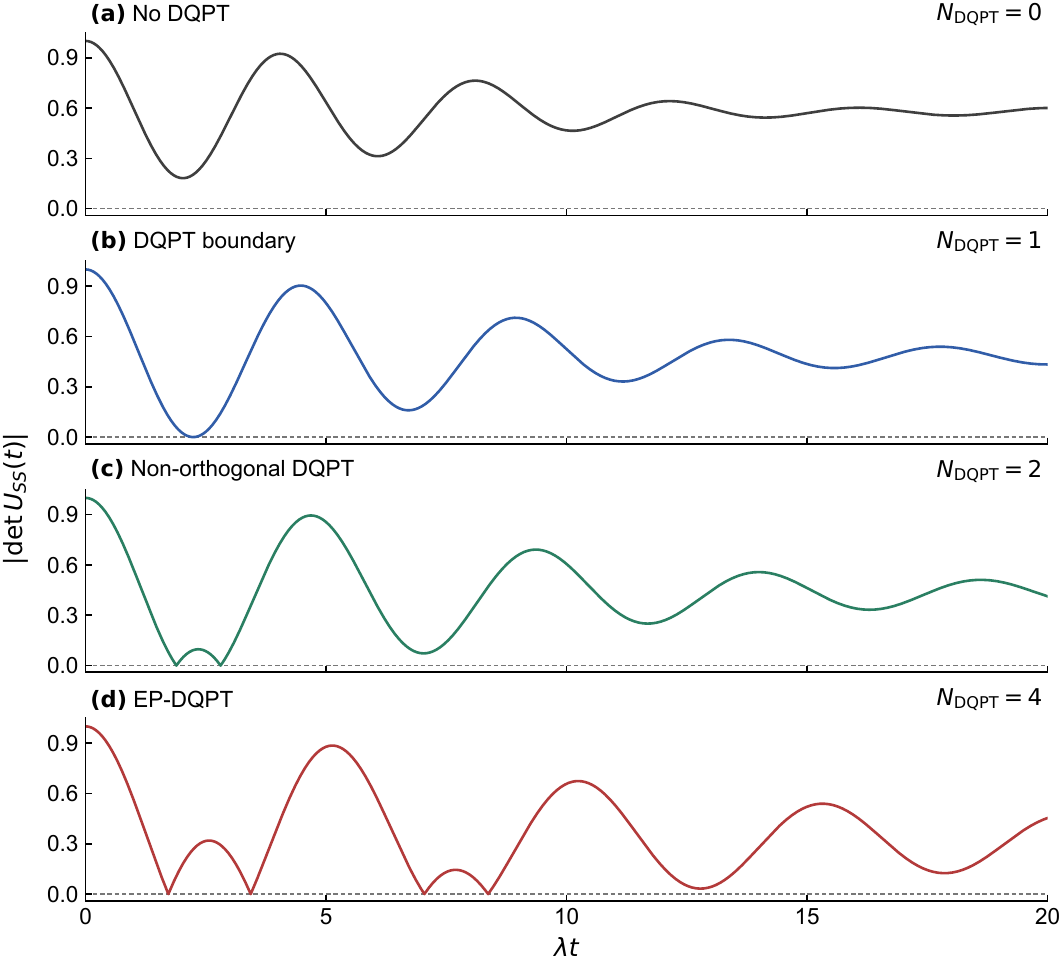}
  \caption{\label{fig:det_dynamics}
    Absolute determinant dynamics, $|\mathcal D(t)|=|\det\US(t)|$, at the four representative points marked in Fig.~\ref{fig:dqpt_phase_dynamics}(a).
    Panels (a)--(d) correspond to no DQPT, the DQPT boundary, a generic non-biorthogonal DQPT, and an EP-DQPT, respectively.
    The dashed line denotes $|\mathcal D(t)|=0$.
    Counting distinct zeros over the phase-diagram observation window, with the quadratic tangency in (b) counted once, gives $N_{\rm DQPT}=(0,1,2,4)$.}
\end{figure}

\refstepcounter{appref}\label{app:det_dynamics}
{\em Appendix B: Time evolution of $\det\US(t)$ and DQPT diagnostics.}---Fig.~\ref{fig:det_dynamics} shows $\mathcal D(t)=\det\US(t)$ at the four representative points of Fig.~\ref{fig:dqpt_phase_dynamics}(a).
The bipartite lattice with real hoppings has a chiral symmetry, $\Gamma H\Gamma=-H$ with $\Gamma=\mathrm{diag}(+1,-1,+1,\dots)$, and $H$ is real symmetric; restricted to $S$, these constraints leave only three real degrees of freedom:
\begin{equation}
  \US(t)=\begin{pmatrix} p & -\ii q\\ -\ii q & r\end{pmatrix},\qquad p,q,r\in\mathbb R,
  \label{eq:USS_real}
\end{equation}
so $\mathcal D=pr+q^2$ and $T\equiv\Tr\US=p+r$ are real functions in real time, which reduces the codimension of the subspace DQPT from two to one and ensures generic real-time crossings without fine-tuning.

Within the plotted window, points A--D contain $N_{\rm DQPT}=(0,1,2,4)$ distinct real critical times, with the quadratic zero at point B counted once.
At point A, $\mathcal D$ remains positive. At point B, it touches zero quadratically, whereas at points C and D it crosses zero.
At point D, only the earliest zero also satisfies $T(t_c)=0$ and realizes the EP-DQPT condition.
The three later zeros at point D are non-EP DQPTs.
The exact equations of motion of the $A$ row, $\dot p=-gq$ and $\dot q=gr$, explain the tangency: an orthogonal zero has $q(t_c)=p(t_c)=0$, hence $\dot{\mathcal D}(t_c)=0$ and $\ddot{\mathcal D}(t_c)=g^2\eta^2>0$, so orthogonal DQPTs are necessarily double zeros with $\alpha=2$.

\refstepcounter{appref}\label{app:normalization}
{\em Appendix C: Normalization and phase conventions.}---For a nonzero rank-one critical propagator $\Uc=\kappa|R_1\rangle\langle L_1|$, we use the following conventions and overlap relations.
(i)~Each state has unit norm in the physical inner product,
\begin{equation}
  \langle R_i|R_i\rangle=\langle L_i|L_i\rangle=1,
  \qquad i=1,2.
  \label{eq:frame_unit_norm}
\end{equation}
(ii)~The right and left null states obey
\begin{equation}
  \langle L_1|R_2\rangle=\langle L_2|R_1\rangle=0.
  \label{eq:frame_cross_orthogonality}
\end{equation}
(iii)~We choose the relative phases within each left--right pair so that
\begin{equation}
  \langle L_1|R_1\rangle=\langle L_2|R_2\rangle=\cos\theta,
  \quad 0\leq\theta\leq\frac{\pi}{2}.
  \label{eq:frame_phase_convention}
\end{equation}
The common overlap is real and nonnegative; its modulus is the dynamical phase rigidity.
In this convention, the eigenvalues are $\eta=\kappa\cos\theta$ and zero, whereas the singular values are $|\kappa|$ and zero.

\refstepcounter{appref}\label{app:thermodynamic_analogy}
{\em Appendix D: Thermodynamic analogy and subspace rate function.}---The analogy with equilibrium phase transitions is direct: the thermodynamic free energy is $F(\beta)=-\beta^{-1}\ln Z(\beta)$, where $Z(\beta)=\Tr\ee^{-\beta H}$.
In DQPT theory, the inverse temperature is continued to a complex variable $z$, and the boundary partition function is $Z(z)=\langle\psi_0|\ee^{-zH}|\psi_0\rangle$~\cite{Heyl2013,Heyl2018}.
On the real-time axis, $z=\ii t$, the boundary partition function becomes the Loschmidt amplitude $\mathcal G(t)$, whose logarithm defines the conventional dynamical free energy up to normalization~\cite{Yin2018AndersonZeros}.
Here we further generalize the single-state amplitude $\mathcal G(t)$ to the subspace determinant $\mathcal D(t)=\det\US(t)$ and define the rate function $f(t)=-\ln|\mathcal D(t)|$.
Every zero $\mathcal D(t_c)=0$ therefore produces a logarithmic divergence of $f(t)$, the subspace counterpart of a thermodynamic free-energy singularity.
Across each simple zero, $\mathcal D$ changes sign and its Pancharatnam phase jumps by $\pi$; no phase jump occurs at the tangential boundary zero.

\refstepcounter{appref}\label{app:higher_order_dqpt}
{\em Appendix E: $m$-level subspaces and $k$th-order DQPTs.}---For $\dim S=m>2$ the criterion $\det\US(t_c)=0$ requires at least one eigenvalue of $\Uc$ to vanish.
If exactly $k$ vanish, we call the transition a $k$th-order subspace DQPT; the critical propagator has rank $m-k$:
\begin{equation}
  \Uc=\sum_{j=1}^{m-k}\kappa_j\,|R_j\rangle\langle L_j|,\qquad\kappa_j\neq0,
  \label{eq:Uc_general_m}
\end{equation}
with every state unit-normalized in the physical norm.
The return amplitude vanishes on two $k$-dimensional families, the kernel states annihilated by $\Uc$ and the states orthogonal to the output span; when the two frames differ, these supply up to $2k$ independent zero-echo directions.
Every non-kernel input is funneled into the $(m-k)$-dimensional output subspace $\mathrm{span}\{|R_1\rangle,\dots,|R_{m-k}\rangle\}$.
For $m>2$, additional state-dependent zeros of the return amplitude can arise from interference within this output subspace.
Such zeros occur even for invertible propagators; for $U=\sigma_x$ one has $\langle0|U|0\rangle=0$ although $\det U=-1$.
They reflect orthogonality between the initial and evolved states rather than rank deficiency, and the determinant criterion excludes them.
This is precisely why Eq.~\eqref{eq:subspace_dqpt_intro}, and not the vanishing of individual echoes, defines the subspace DQPT.

\refstepcounter{appref}\label{app:quasiparticle_dqpt}
{\em Appendix F: Interaction-induced quasiparticle DQPT.}---For an interacting many-body system, no subspace needs to be chosen by hand: the single-particle sector itself plays that role.
The retarded Green's function restricted to a set of single-particle orbitals $\{|n\rangle\}$ is
\begin{equation}
  G^R_{mn}(t) = -\ii\,\theta(t)\,\langle\Omega|\{c_m(t),\, c_n^\dagger\}|\Omega\rangle,
  \label{eq:GR_def}
\end{equation}
with $|\Omega\rangle$ the many-body ground state and $c_m(t)=\ee^{\ii Ht}c_m\ee^{-\ii Ht}$.
It obeys the same Dyson structure as Eq.~\eqref{eq:USS_resolvent}, with the self-energy $\Sigma^R(\omega)$ now generated by interactions.
In the noninteracting limit $\Sigma^R\to0$, one has $G^R_{mn}(t)=-\ii\,\theta(t)[\ee^{-\ii ht}]_{mn}$ with $h$ the single-particle Hamiltonian.
Then $\det G^R(t)\propto\ee^{-\ii t\Tr h}$ never vanishes, consistent with the no-go theorem of Appendix~\ref{app:self_energy_no_go}.
Interactions can therefore drive $\det G^R(t_c)=0$, defining a quasiparticle DQPT that is entirely interaction induced.
The rank-one structure of the critical propagator and the non-biorthogonal Born rule then carry over to the quasiparticle sector unchanged.

\refstepcounter{appref}\label{app:memory_kernel}
{\em Appendix G: Explicit memory kernel and the narrow-band regime.}---The finite bandwidth of the environment produces a memory kernel whose timescale increases as $J_c$ decreases.
Let $|\psi_S(t)\rangle$ and $|\psi_E(t)\rangle$ denote the subsystem and environment amplitudes, with $|\psi_E(0)\rangle=0$.
Solving the environment equation gives
\begin{equation}
  |\psi_E(t)\rangle=-\ii\lambda\int_0^t\mathrm{d}s\,
  \ee^{-\ii H_E(t-s)}|1\rangle\langle B|\psi_S(s)\rangle.
  \label{eq:environment_elimination}
\end{equation}
Substitution into the subsystem equation yields
\begin{equation}
  \begin{aligned}
    \ii\partial_t|\psi_S(t)\rangle
    &=H_S|\psi_S(t)\rangle\\
    &\quad-\ii\int_0^t\mathrm{d}s\,K(t-s)
    |B\rangle\langle B|\psi_S(s)\rangle,
  \end{aligned}
  \label{eq:memory_equation}
\end{equation}
where the semi-infinite chain has the exact surface kernel
\begin{equation}
  \begin{aligned}
    K(\tau)&=\lambda^2\langle1|\ee^{-\ii H_E\tau}|1\rangle\\
    &=\lambda^2\frac{J_1(2J_c\tau)}{J_c\tau},
    \qquad K(0)=\lambda^2,
  \end{aligned}
  \label{eq:surface_memory_kernel}
\end{equation}
with $J_1$ the Bessel function of the first kind.
The dependence on $J_c\tau$ sets an environmental memory timescale $\tau_E\sim J_c^{-1}$.
Relative to the coupling timescale $\tau_\lambda=\lambda^{-1}$, one has $\tau_E/\tau_\lambda\sim\lambda/J_c$.
Thus, for $J_c<\lambda$, the kernel retains the subsystem's history over the coupling dynamics, explaining the pronounced non-Markovian features in the narrow-band regime.

In the wide-band limit $J_c\to\infty$ at fixed $g$ and $\Gamma\equiv\lambda^2/J_c$, the kernel becomes localized at $\tau=0$, with $\int_0^\infty K(\tau)\,\mathrm{d}\tau=\Gamma$.
For $t>0$, Eq.~\eqref{eq:memory_equation} then reduces to
\begin{equation}
  \ii\partial_t|\psi_S(t)\rangle
  =\bigl(H_S-\ii\Gamma|B\rangle\langle B|\bigr)
  |\psi_S(t)\rangle.
  \label{eq:memory_wide_band}
\end{equation}
This is the constant instantaneous self-energy of Appendix~\ref{app:self_energy_no_go}, giving $\det\US(t)=\ee^{-\Gamma t}$ and hence no finite-time subspace DQPT.

\clearpage
\onecolumngrid
\section*{Supplemental Material}
\twocolumngrid

{\em Orthogonality of the flat-band-connected boundary.}---We first solve the flat-band limit exactly and then continue its DQPT boundary to finite $J_c$.
In the flat-band limit $J_c=0$, the environment reduces to one site and the model becomes a three-level system.
With $\US(t)=\left(\begin{smallmatrix}p&-\ii q\\-\ii q&r\end{smallmatrix}\right)$ and $\Omega=\sqrt{g^2+\lambda^2}$, the amplitudes are
\begin{equation}
  \begin{aligned}
    p_0(t) & =\frac{\lambda^2+g^2\cos(\Omega t)}{\Omega^2}, \\
    q_0(t) & =\frac{g}{\Omega}\sin(\Omega t),               \\
    r_0(t) & =\cos(\Omega t).
  \end{aligned}
  \label{eq:flat_band_amplitudes}
\end{equation}
The corresponding determinant is
\begin{equation}
  \mathcal D_0(t)=p_0r_0+q_0^2
  =\frac{g^2+\lambda^2\cos(\Omega t)}{\Omega^2}.
  \label{eq:flat_band_determinant}
\end{equation}
Its first DQPT boundary occurs at
\begin{equation}
  g_0=\lambda,\qquad t_0=\frac{\pi}{\sqrt{2}\lambda},
  \label{eq:flat_band_boundary}
\end{equation}
where $p_0(t_0)=q_0(t_0)=0$ and $r_0(t_0)=-1$.
Therefore
\begin{equation}
  \US(t_0)=\begin{pmatrix}0&0\\0&-1\end{pmatrix}
  =\US^{\dagger}(t_0),
  \label{eq:flat_band_boundary_propagator}
\end{equation}
so the flat-band boundary is an orthogonal DQPT.

We next continue this boundary to finite $J_c$.
The equations $p(g,t,J_c)=0$ and $q(g,t,J_c)=0$ have a nondegenerate solution at $(g_0,t_0,0)$ because
\begin{equation}
  \left.\det\frac{\partial(p,q)}{\partial(g,t)}\right|_{(g_0,t_0,0)}=1.
  \label{eq:pq_jacobian}
\end{equation}
The implicit-function theorem therefore gives a unique local analytic branch $g_{\rm o}(J_c),t_{\rm o}(J_c)$ satisfying $p=q=0$.
Along this branch, $\mathcal D=pr+q^2=0$ and $\dot{\mathcal D}=0$.
At the flat-band point, the DQPT boundary equations are also nondegenerate,
\begin{equation}
  \left.\det\frac{\partial(\mathcal D,\dot{\mathcal D})}{\partial(g,t)}\right|_{(g_0,t_0,0)}=\lambda\neq0,
  \label{eq:fold_jacobian}
\end{equation}
so local uniqueness identifies the $p=q=0$ branch with the DQPT boundary.

The gray curve in Fig.~\ref{fig:dqpt_phase_dynamics}(a) is determined directly from the DQPT count: it separates the $N_{\rm DQPT}=0$ and $N_{\rm DQPT}>0$ regions.
Throughout the plotted range, we find numerically that this independently determined DQPT boundary coincides with the nondegenerate $p=q=0$ branch continued from $J_c=0$.
No branch exchange is encountered, so the boundary shown in Fig.~\ref{fig:dqpt_phase_dynamics}(a) remains on this branch and gives
\begin{equation}
  \Uc=\begin{pmatrix}0&0\\0&r(t_c)\end{pmatrix}=\Uc^{\dagger}.
  \label{eq:continued_boundary_propagator}
\end{equation}
Hence the DQPT boundary shown in Fig.~\ref{fig:dqpt_phase_dynamics}(a) is normal and has an orthonormal eigenframe.

\end{document}